\documentclass[10pt,journal,compsoc]{IEEEtran}
\usepackage{graphics}
\usepackage{graphicx}
\usepackage{algorithm}  
\usepackage{algpseudocode}  
\usepackage{amsmath}  
\usepackage{amsfonts}
\usepackage{amssymb}
\usepackage{latexsym}
\usepackage{caption, subcaption}
\usepackage{ulem}
\usepackage{booktabs}
\usepackage{listings}
\usepackage{epstopdf}
\usepackage{chngpage}
\usepackage{diagbox} 
\usepackage{adjustbox}
\usepackage{setspace}
\usepackage{multirow}
\usepackage{CJK}
\usepackage{ulem}

\usepackage[utf8]{inputenc} 
\usepackage[T1]{fontenc} 
\usepackage{fouriernc} 

\ifCLASSOPTIONcompsoc
  \usepackage{cite}
\else
  \usepackage{cite}
\fi
\ifCLASSINFOpdf
\else
\fi
\begin{document}
%
\title{\huge{Effective Scaling of Blockchain Beyond Consensus Innovations and Moore's Law}}
%
%
%
%

\author{Yinqiu~Liu,
        Kai~Qian,
        Jianli~Chen,
        Kun~Wang,
and     Lei~He

\IEEEcompsocitemizethanks{\IEEEcompsocthanksitem Y. Liu, K. Qian, J. Chen, K. Wang, and L. He are with the Department of Electrical and Computer Engineering, University of California, Los Angeles, CA, USA.
\IEEEcompsocthanksitem Corresponding authors: Kun Wang (e-mail: wangk@ucla.edu) and Lei He (e-mail: lhe@ee.ucla.edu).
}}

%
%

\markboth{}%
{Shell \MakeLowercase{\textit{et al.}}: Bare Demo of IEEEtran.cls for Computer Society Journals}
%



\IEEEtitleabstractindextext{%
\vspace{-0.5cm}
\begin{abstract}
As an emerging technology, blockchain has achieved great success in numerous application scenarios, from intelligent healthcare to smart cities. 
However, a long-standing bottleneck hindering its further development is the massive resource consumption attributed to the distributed storage and consensus mechanisms. 
This makes blockchain suffer from insufficient performance and poor scalability. 
Here, we analyze the recent blockchain techniques and demonstrate that the potential of widely-adopted consensus-based scaling is seriously limited, especially in the current era when Moore's law-based hardware scaling is about to end. 
We achieve this by developing an open-source benchmarking tool, called \textsf{Prism}, for investigating the key factors causing low resource efficiency and then discuss various topology and hardware innovations which could help to scale up blockchain. 
To the best of our knowledge, this is the first in-depth study that explores the next-generation scaling strategies by conducting large-scale and comprehensive benchmarking.
\end{abstract}

}

\maketitle

\IEEEdisplaynontitleabstractindextext

%
\IEEEpeerreviewmaketitle

\IEEEraisesectionheading{\section{Introduction}\label{sec:introduction}}

%
%
%
%
\vspace{-0.1cm}
Since Satoshi Nakamoto introduced the electronic cash named "Bitcoin", blockchain technology has experienced rapid development, with a market worth more than 263 US\$ billion \cite{BTC2019}. 
Apart from payments, it becomes instrumental in various academic and industrial fields, such as Internet of Things (IoT) \cite{8605485B, 8661654A}, Artificial Intelligence \cite{Jacobsen1B, 7995134}, and big data \cite{7997A, 7999A2}. 
In the database context, blockchain refers to the append-only digital ledgers which consist of chronologically-ordered blocks. 
The blocks that carry numerous certified transactions transferring value or information, are connected one by one, then form a directed chain. 
In this way, traceable recording services with high-level data integrity can be enabled for almost every application scenario.

To avoid the malicious tampering in public environments, distributed operating manners are adopted. 
Specifically, all participants synchronize and store the chain states in parallel via consensus mechanisms and gossiping protocols, thereby constructing a peer-to-peer (P2P) network, that is, blockchain network \cite{8760539A}. 
As there is no reliance on certificate authorities (CAs), blockchain significantly improves the decentralization and security. 
However, such accomplishments need to be supported by tremendous resources and energy \cite{8000G}. 
Viewing blockchain as a global computer, Fig. 1 shows how the computation and storage overhead of popular blockchain networks have changed over the past ten years, and clearly illustrates an exponential growth \cite{chart1}.
Moreover, the ever-increasing resource consumption contributes little to promote blockchain's performance of processing transactions. 
In fact, massive resources are invested in ensuring security or strengthening robustness. 
As the field develops, the insufficient performance becomes a bottleneck hindering the further blockchain deployments in the scenarios where peers are resource-constrained or with high workloads.

To address the above issues, researchers have begun exploring techniques to optimize blockchain in terms of performance. 
The ultimate judgment of such optimizations is based on the scalability, defined as the capability to expand P2P network under the premise of sustaining enough performance \cite{Scalability}. 
In other words, the performance of blockchain network should be linearly and positively correlated with its scale (measured by full node number and/or total available resources), or should at least remain stable. 
Consensus innovations, focusing on designing lightweight consensus workflows, provide a relatively easy path towards scaling. 
However, researchers encounter roadblocks when intending to scale up blockchain by orders of magnitude --- the resource efficiency still remains in situ. 
Simultaneously, Moore's law-based hardware evolution, from central processing units (CPUs) to graphics processing units (GPUs), field-programmable gate arrays (FPGAs), and application specific integrated circuits (ASICs), seems to mainly intensify the energy consumption rather than providing scalability optimizations \cite{8048662, QC}. 
Effectively scaling up blockchain, without affecting its properties of decentralization and security, constructs a "trilemma" and poses daunting challenges for researchers \cite{trilemma}.
\begin{figure*}[htbp]
\vspace{-0.1cm}
\centering
\begin{minipage}{\textwidth}
\centering
    \begin{minipage}{0.02\textwidth}
	\includegraphics[width=0.9\textwidth]{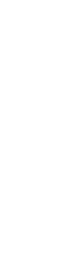}
	\end{minipage}
	\begin{minipage}{0.45\textwidth}
	\includegraphics[width=0.9\textwidth]{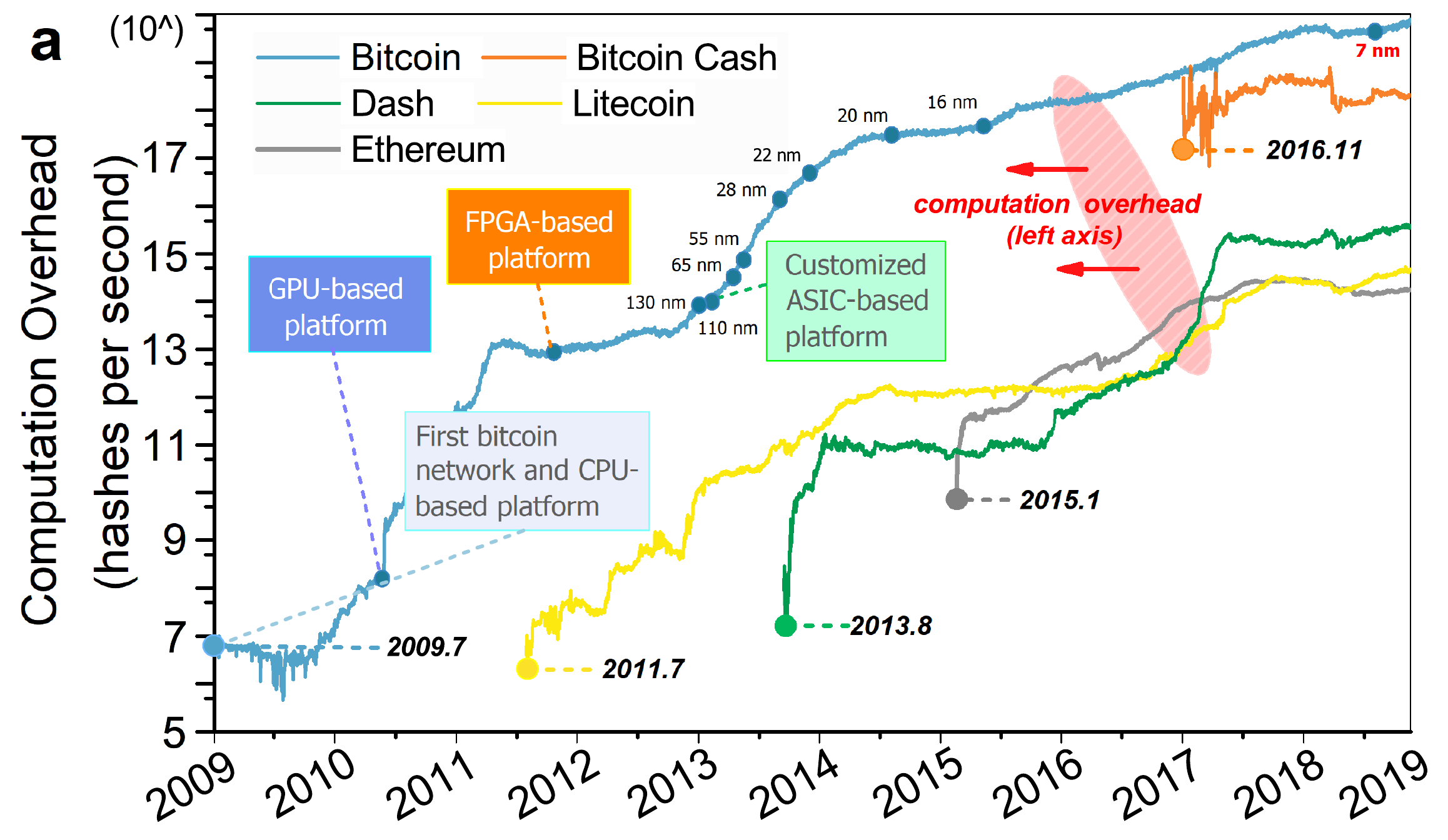}
	\end{minipage}
	\begin{minipage}{0.01\textwidth}
	\includegraphics[width=0.9\textwidth]{fill.png}
	\end{minipage}
	\begin{minipage}{0.45\textwidth}
	\includegraphics[width=0.9\textwidth]{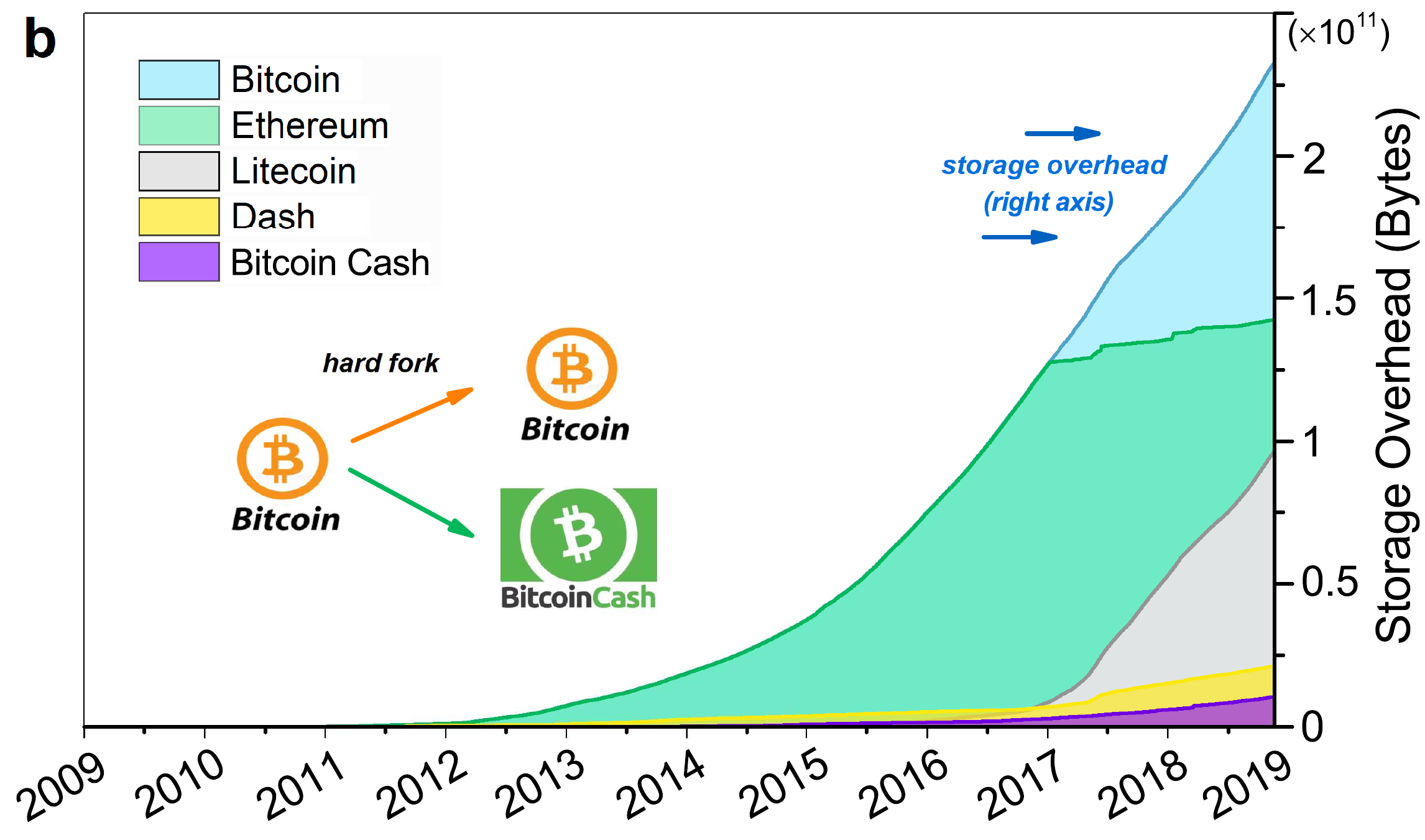}
	\end{minipage}
	\label{fig:computational-cost}
	\caption{\textbf{Resource overhead of well-known blockchain networks} \textbf{a,}  \textit{The computation overhead and Moore's law-based hardware scaling in Bitcoin network, 2009-2019. The left y axis is in log scale.} \textbf{b,} \textit{The storage overhead.} \textit{Note that Bitcoin Cash originates from Bitcoin, so their inital storage overhead is the same. Data taken from \cite{chart1} and \cite{8048662}.}}
\end{minipage}
\end{figure*}

In this Perspective, we analyse the scalability of state-of-the-art blockchain designs and demonstrate that deficiencies exist both at their current performance and scaling rates. 
Furthermore, we demonstrate that the potential of consensus-based scaling is limited --- it is bordering its upper bound of scaling ability. 
As scalability is tightly related to the resource consumpution, we develop a benchmarking tool named \textsf{Prism} (\textit{https://github.com/Lancelot1998/Prism}) to investigate the hidden factors hindering blockchain techniques. 
Looking to the future, we suggest that innovations from two aspects could help to scale up blockchain: topology-based scaling and hardware-assisted scaling.

\section{Deficiencies for the Consensus-based Scaling in the Past Decade}\label{sec:deficiency}
To precisely describe the history and effectiveness of 10 years of consensus-based scaling, we first quantitatively examine the performance (including throughput and latency) of mainstream blockchain designs with vital consensus innovations. 
Additionally, we analyse the scalability of these designs, as well as that of representative academic proposals. 

Using identity management strategy, we divide blockchain into public- and private-chain \cite{8246573A}. 
The former constructs the anonymous and permissionless networks over globally distributed peers, with considerable hardware heterogeneity. 
Consequently, their performance benchmarking can only be organized by official groups, whose reports are publicly available on blockchain exporters.
In contrast, for private-chain applied in enterprise-level scenarios, several permissioned peers manipulated by certain departments form an autonomous organization. 
Due to the absence of a worldwide public network, a testbed is needed before benchmarking private-chain designs. To this end, we set up a standard testbed, whose configurations represent the practical cases of private-chain applications. 
In detail, our autonomous organization possesses 10 geo-distributed servers with fixed resources (2-core Intel Xeon CPU, 8GB RAM, 40GB hard drive, and 5Mbps bandwidth). 
The selected benchmarking techniques include Caliper \cite{Caliper}, BLOCKBENCH \cite{Dinh:2017:BFA:3035918.3064033B}, and \textsf{Prism}, with which we cover most of the mainstream private-chain designs. 
Note that \textsf{Prism} is a self-designed benchmarking tool and acquires higher accuracy and compatibility than others with the help of docker technology.

To the best of our knowledge, this Perspective is the first to conduct benchmarking for various private-chain designs in a uniform testbed, making the performance comparison meaningful and informative. 
We are particularly interested in whether consensus-based scaling could improve the performance following application requirements and effectively scale up blockchain. 
\begin{figure*}[htb]
	\centering
	\scalebox{1}[1]{
	\includegraphics[width=13.35cm, height=6cm]{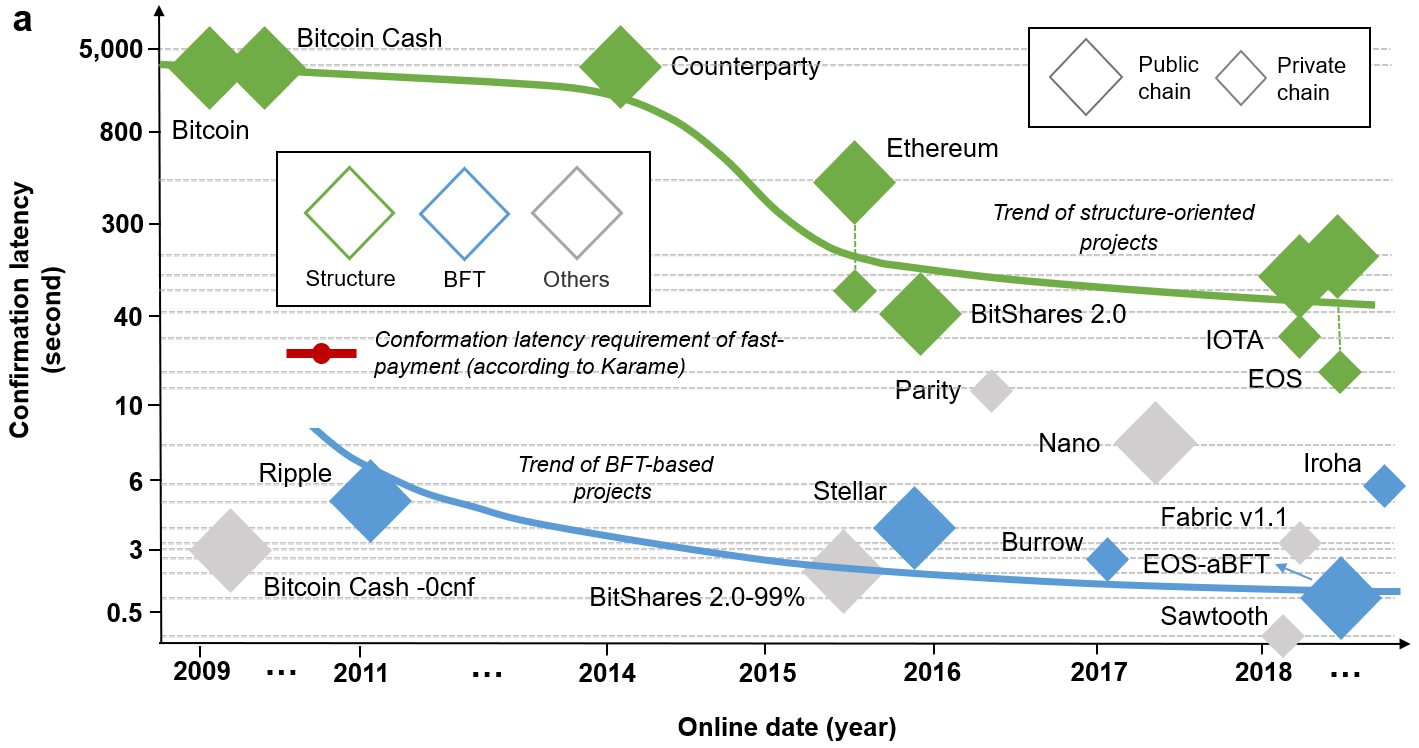}}
	\label{fig:transaction}
\end{figure*}
\vspace{0.5cm}
\begin{figure*}[htb]
	\centering
	\scalebox{1}[1]{
	\includegraphics[width=13.35cm, height=6cm]{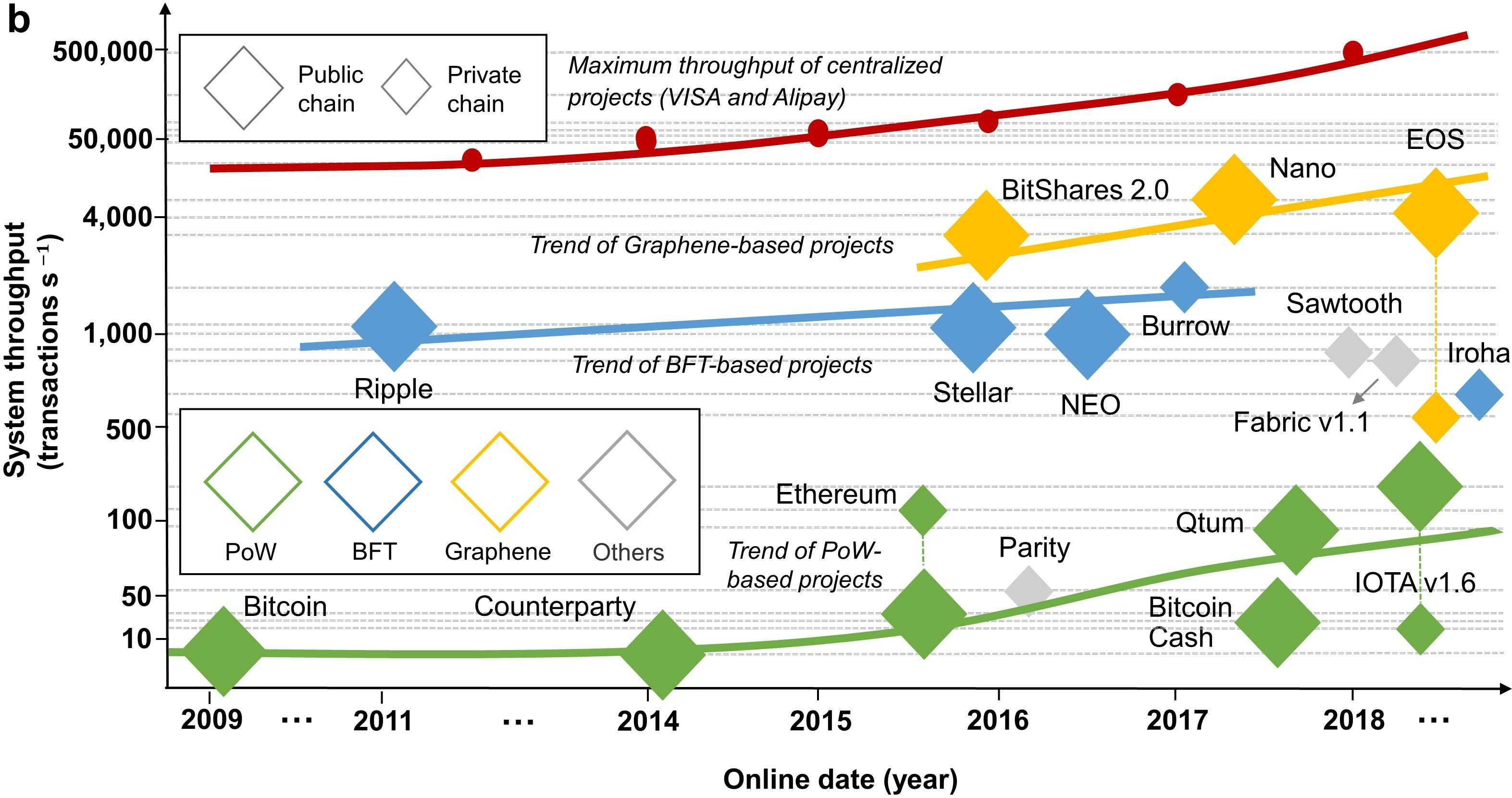}}
	\caption{\textbf{Performance deficiency existing in mainstream blockchain designs and the potential of consensus-based scaling.} \textbf{a,} \textit{The average confirmation latency demand by leading blockchain designs against year.} \textbf{b,} \textit{The peak throughput of leading blockchain designs against year. Note that these sequential proposals reflect the history of consensus-based scaling since their performance optimizations are mainly implemented based on consensus innovations. Conversely, innovations from other perspectives are still immature, especially lacking practical benchmarks. Data arranged in \textsf{Prism}.}}
	\label{fig:transaction}
\vspace{-0.1cm}
\end{figure*}

\vspace{-0.6cm}
\subsection{Performance deficiency}
In Fig. 2a, we illustrate the confirmation latency (that is, the time interval for irreversibly confirming one transaction) of the leading blockchain designs. 
Considering that the bases used to judge "irreversibility" have differences, we classify the involved designs into two major categories: structure-oriented and Byzantine Fault Tolerance (BFT)-based ones, both of which contain plentiful derivatives. 

The structure-oriented blockchain appeared first and stemmed from Bitcoin, which confirms transactions via the Nakamoto consensus.
In this case, local ledger extends linearly and only the transactions on the longest chain among multiple conflicting forks will be confirmed. 
To guarantee the irreversibility of their transaction confirmations in such a competitive environment, clients wait for the proof of 6 subsequent blocks \cite{Gilad:2017:ASB:3132747.3132757A}. 
Given that each block creation roughly takes 600s, the latency fluctuates around 3600s. 
Since Bitcoin Cash \cite{BCH} and Counterparty \cite{CPT} share an equivalent architecture with Bitcoin, the confirmation latency has remained constant for several years. 
In 2015, Ethereum was developed, using Greedy Heaviest-Observed Sub-Tree (GHOST) protocol to accommodate an increased block-creation rate (15s per block) \cite{GHOSTA}. 
Accordingly, chain structure evolved from linearity to parental tree and the capability of forks was exploited.
Then, the confirmation latency dramatically declined to less than 570s. 
BitShares 2.0 and EOS, emerging between 2015 and 2018, substituted permissionless competitions for geo-ordered block creations by permissioned block producers (BPs) \cite{EOS-IO}. 
Reduced amount of competitors and pre-configured consensus workflows are conducive to accelerating the process of reaching irreversibility. 
Thus, confirmations in BitShares 2.0 and EOS cost approximately 45s and 198s, respectively, outperforming the predecessors again. 
Since then, the improvement on latency stalled, even though IOTA further upgraded tree-structured ledger to a Directed Acyclic Graph (DAG) \cite{DAGA}. 
This is due to the fact that the structure-oriented designs view ledger extensions as the confirmation process, which is time-consuming and cannot be thoroughly alleviated through consensus innovations.

Under BFT-based designs, pending transactions call for commitments from no less than 2\textit{f}\,+\,1 peers (replicas), where \textit{f} represents the sum of attackers in P2P network \cite{Castro:1999:PBF:296806.296824A}.
In Fig. 4a, we observe that the confirmation latency of BFT-based designs varies between 3-10s, with a slight drop over time. 

Note that the confidence coefficient of "irreversibility" also influences clients' judgements. 
If the submitted transaction only carries micro currency, appropriately relaxing the confidence interval could  simplify the confirmation process without a risk of serious losses, for instance, zero-confirmation for Bitcoin. 
With respect to the standard of latency, Karame \textit{et al}. \cite{Karame:2012:DFP:2382196.2382292A} stressed that the validity of each pending transaction would better be determined within 30s, termed as fast payments. 
Although it seems that BFT-based designs have outperformed such a baseline, workloads of novel applications sparked by advanced techniques increase dramatically, leaving ever-decreasing time for confirmations. 
For example, the average latency of 5G-driven applications is less than 100ms \cite{5GA}. 
Moreover, centralized counterparties, like Alipay, have already achieved instant confirmation. 
All the designs involved in Fig. 2a use the strategy of consensus-based scaling, which is the most widely-adopted scaling method.  
As both trends tend to slow down, we conclude that deficiency exists with respect to the confirmation latency.

Similarly, Fig. 2b depicts the trends of peak throughput (that is, the maximum ability of processing concurrent transactions per second), following the leading blockchain designs. 
Note that we compare the block creation strategies during the consensus workflows and thus classify the above designs into different categories. 

From 2009 to 2015, Bitcoin and its imitators (called Altcoins) dominated the market and their consensus mechanisms are diverse versions of Proof-of-Work (PoW). 
Under PoW, the validity of blocks depends on creators' workload of hash collisions trying to figure out tough cryptography puzzles, named mining \cite{8735815A}. 
Since all miners in the P2P network simultaneously commit massive resources into mining while only one can win the competition and create a block in each round, PoW is characterized by huge resource waste and low performance.
Even though subsequent researchers revised standard PoW, such as by adopting the indicator "stake" to mitigate the computing power usage, thus proposing so-called hybrid consensus, PoW-based designs still hold weak throughput \cite{HybridA, pass_et_al:LIPIcs:2017:8004A} (from 3.3-6.67 TPS for Bitcoin to about 250 TPS for IOTA).

To reduce the resource overhead, researchers have carried out relentless efforts for redesigning lightweight consensus. 
BFT, initially aiming to achieve fault tolerance in distributed systems, conducts byzantine-style voting to create blocks. 
Since such process achieve higher time- and resource-efficiency, the peak throughput of BFT-based designs enters in the 1000 TPS-1500 TPS range. 

\vspace{-0.05cm}
After 2015, Graphene aroused widespread interest because this framework drastically stimulated the throughput to more than 3500 TPS through employing BP. 
For example, BPs of EOS take turns to create one batch containing 6 micro-blocks, after which they broadcast the proposals to neighbouring BPs within 3s \cite{EOS-IO}. 
Graphene-based designs are seriously questionable in terms of decentralization and security due to the prior configurations of permissions, geographic orders, even block-creation rates. 
That being said, authoritative BPs make EOS almost reach the theoretical upper bounds of throughput for current blockchain techniques, where the transaction processing speed over the whole P2P network is only capped at that of BPs. 
However, the overall throughput of blockchain still displays a gigantic, widening gap compared to the application demand or centralized counterparties. 
As shown in Fig. 2b, Alipay and Visa both bypassed 50000 TPS around 2015 \cite{Visa}. 
Through strengthening datacenters, this indicator is growing by a factor of two per year. 
Conversely, consensus-based scaling no longer offers help, as all three trends flatten out.
Given that innovations from other perspectives, such as multi-chain topology, are still in their infancy, it is clear that deficiency also exists in throughput.
\begin{figure*}[htpb]
	\vspace{1pt}
	\centering
	\scalebox{1}[1]{
	\includegraphics[width=14cm, height=6.6cm]{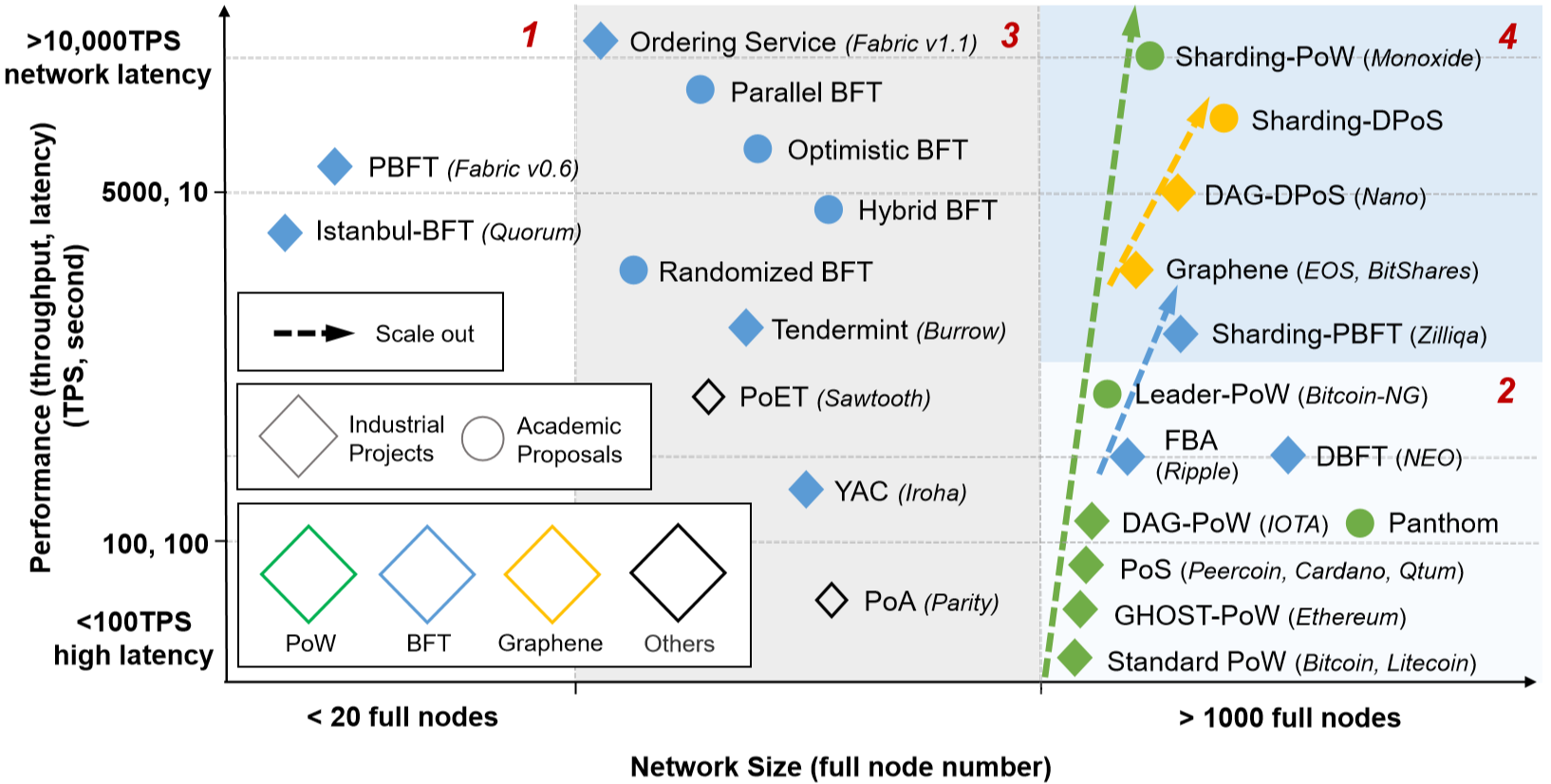}}
	\vspace{0.1cm}
	\caption{\textbf{Scalability deficiency of mainstream blockchain designs and the validity of typical scaling proposals.}  \textit{In this figure, the diamond and circular items represent the consensus mechanisms from industrial projects and academic scaling proposals, respectively. The bracketed content of each item lists well-proven blockchain designs based on it. Furthermore, three dotted lines show the scaling trends of classic consensus families. Apparently, scalability breakthroughs are not attributed to lightweight consensus mechanisms while are stimulated by topology innovations (like sharding of Monoxide and Zilliqa, or BPs of EOS). Data taken from \cite{studyA} and \cite{227661B}.}}
	\label{fig:transaction}
\end{figure*}

The insights from Fig. 2 demonstrate that 10 years of consensus-based scaling fails to address the skyrocketing demand on performance. 
We focus on latency and throughput only because they are the most important performance metrics for blockchain.
Recently, Facebook presented its financial infrastructure named Libra \cite{Libra}, whose consensus mechanism is Libra-BFT. 
Since the announced throughput of Libra-BFT (1000 TPS) is in line with our predictions for BFT-based designs, the accuracy of Fig. 2 gets further verified.

\vspace{0.1cm}
\subsection{Scalability deficiency}
As we mentioned before, scalability reflects the performance tendency as more participants join in the blockchain network. 
To visualize the scalability of mainstream blockchain designs, Fig. 3 shows their throughput and latency versus the number of full nodes they can support at most. 
In addition, we take the validity of well-known scaling proposals, primarily those which underwent practical benchmarks, into consideration. 
In Fig. 3, we first analyse the items located in $\textbf{\textit{Areas}}\; \textbf{\textit{1-3}}$. 

Practical BFT (PBFT)-based designs within $\textbf{\textit{Area}}\; \textbf{\textit{1}}$ excel in performance. 
In theoretical cases, they obtain a throughput over 10000 TPS and a latency only bounded by bandwidth. 
However, PBFT splits the consensus workflows into three successive phases, each of which executes network-wide multicast to collect commitments. 
For the P2P network composed of $n$ full nodes, the communication complexity required for one round of ledger extension is $O(n^2)$ \cite{Castro:1999:PBF:296806.296824A}. 
Despite an outstanding throughput in small-scale networks, such superiority drastically decays as more full nodes participate. 
Hence, the adaptability of PBFT-based blockchain is severely undermined, in which the P2P network can only accommodate dozens of (in \cite{ENERGY1}, 16) peers. 

Conversely, the throughput of PoW-based items marked in $\textbf{\textit{Area}}\; \textbf{\textit{2}}$ is generally less than 2000 TPS and the latency keeps in the order of minutes. 
Although the huge resource consumption greatly affects the performance, this set of blockchain is capable of retaining stability when the network expands to thousands of full nodes. 
The corresponding communication complexity is just $O(n)$, or even $O(1)$ \cite{8760539A}. 
Such robustness enables blockchain to meet the requirements of not only small-scale but also globally-distributed deployments. 

Some proposals within $\textbf{\textit{Area}}\; \textbf{\textit{3}}$, mostly belonging to PBFT's variations, seem to better balance between the performance and network scale. 
Unfortunately, the majority of them are short of practical benchmarks in large-scale network environments and thus their scalability is subject to further research \cite{studyA}. 
Clearly, the blockchain within $\textbf{\textit{Areas}}\; \textbf{\textit{1-3}}$ has distinct deficiencies in this metric.

Following three scaling trends in Fig. 3, we observe that the items within $\textbf{\textit{Area}}\; \textbf{\textit{4}}$ obtain satisfying optimizations in regard to both performance and the network scale. 
Unlike the aforementioned techniques whose scaling relies on consensus innovations, the items in this area adopt well-studied consensus mechanisms, such as PoW and PBFT, while rebuilding the network topology.
In particular, participants are clustered for maintaining multiple chains in parallel, instead of placing resources into one single chain. 
In this way, the large-scale network can be parallelized into smaller entities (side-chain or shard), which independently allocate resources \cite{8539529A}. 
With the cross-chain/shard communication mechanisms, peers from different entities are still accessible to exchange information. 
In this solution, performance of the entire blockchain network can rise positively with the entity number. 
For instance, Zilliqa clustered 600 peers in one shard then reached [1218, 1752, 2488] TPS in the P2P network with a scale of [1800, 2400, 3600] full nodes, respectively \cite{Zilliqa}. 
We reckon that topology-based scaling is able to overcome the roadblocks that impede current consensus-based scaling and provide a novel path towards effectively scaling up blockchain in the future. 
\begin{figure*}[htpb]
\vspace{0.2cm}
\centering
\begin{minipage}{1\textwidth}
	\centering
	\begin{minipage}{0.26\textwidth}
	\includegraphics[width=\textwidth]{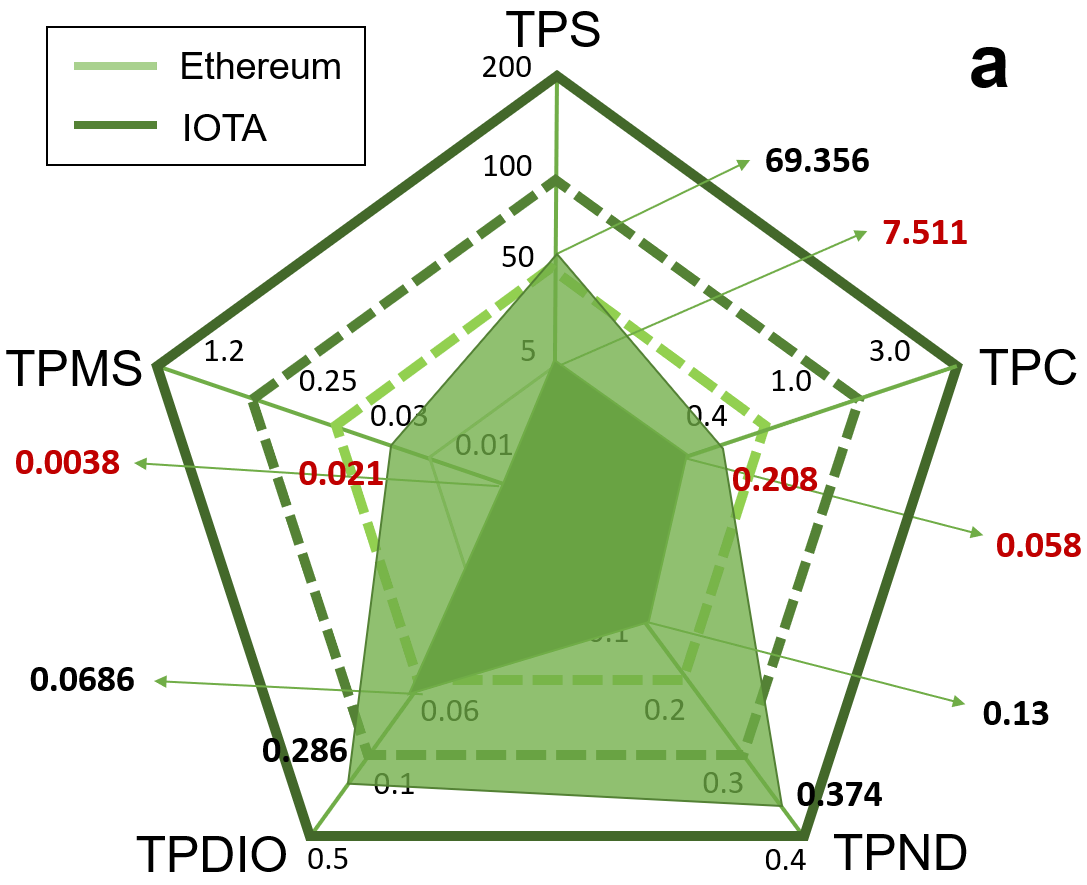}
	\end{minipage}
	\begin{minipage}{0.01\textwidth}
	\includegraphics[width=\textwidth]{fill.png}
	\end{minipage}
	\begin{minipage}{0.26\textwidth}
	\includegraphics[width=\textwidth]{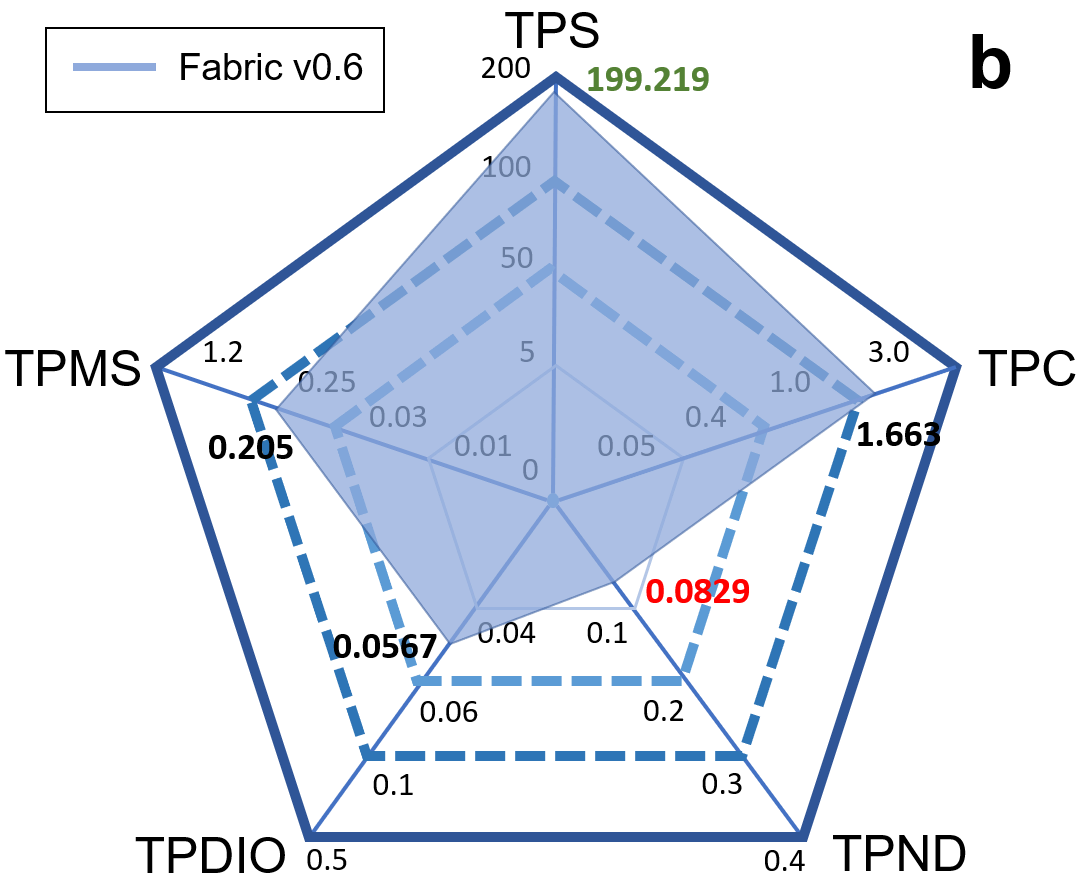}
	\end{minipage}
	\begin{minipage}{0.01\textwidth}
	\includegraphics[width=\textwidth]{fill.png}
	\end{minipage}
	\begin{minipage}{0.26\textwidth}
	\includegraphics[width=\textwidth]{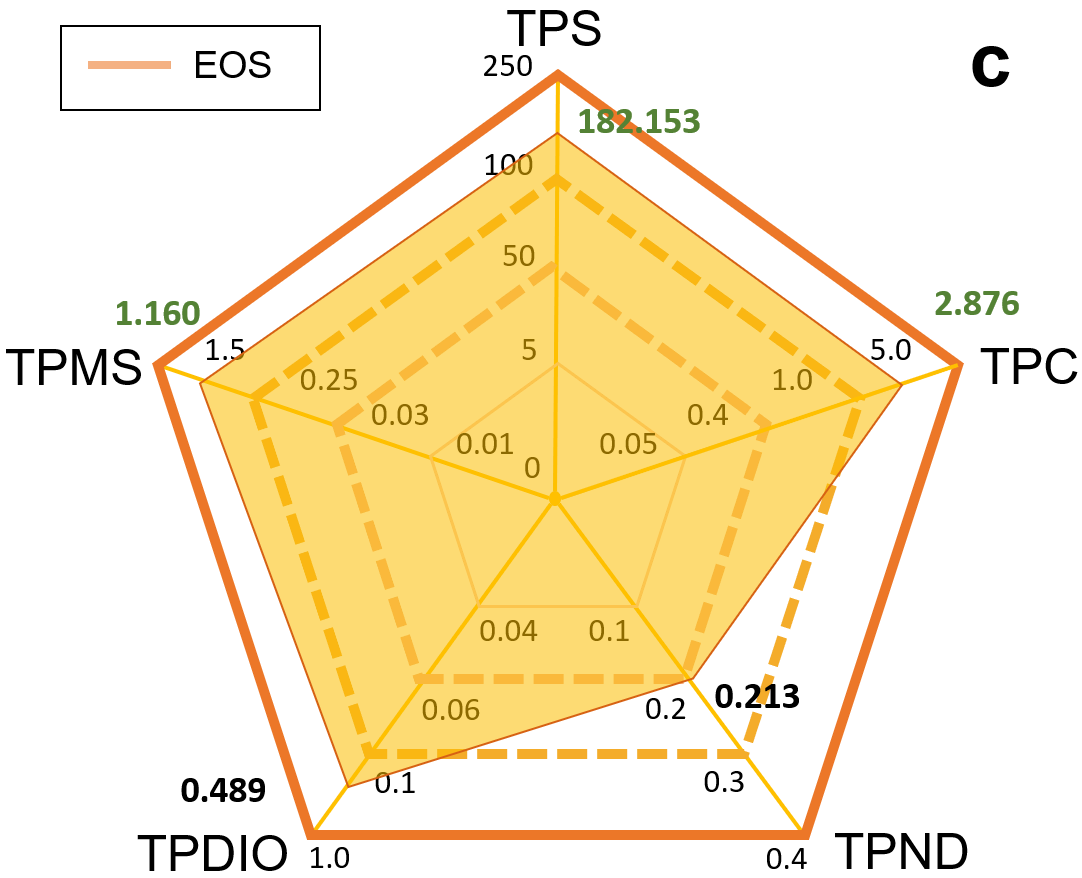}
	\end{minipage}
\end{minipage}
\end{figure*}
\begin{figure*}[htpb]
\vspace{0.2cm}
\centering
\begin{minipage}{1\textwidth}
	\centering
	\begin{minipage}{0.26\textwidth}
	\includegraphics[width=\textwidth]{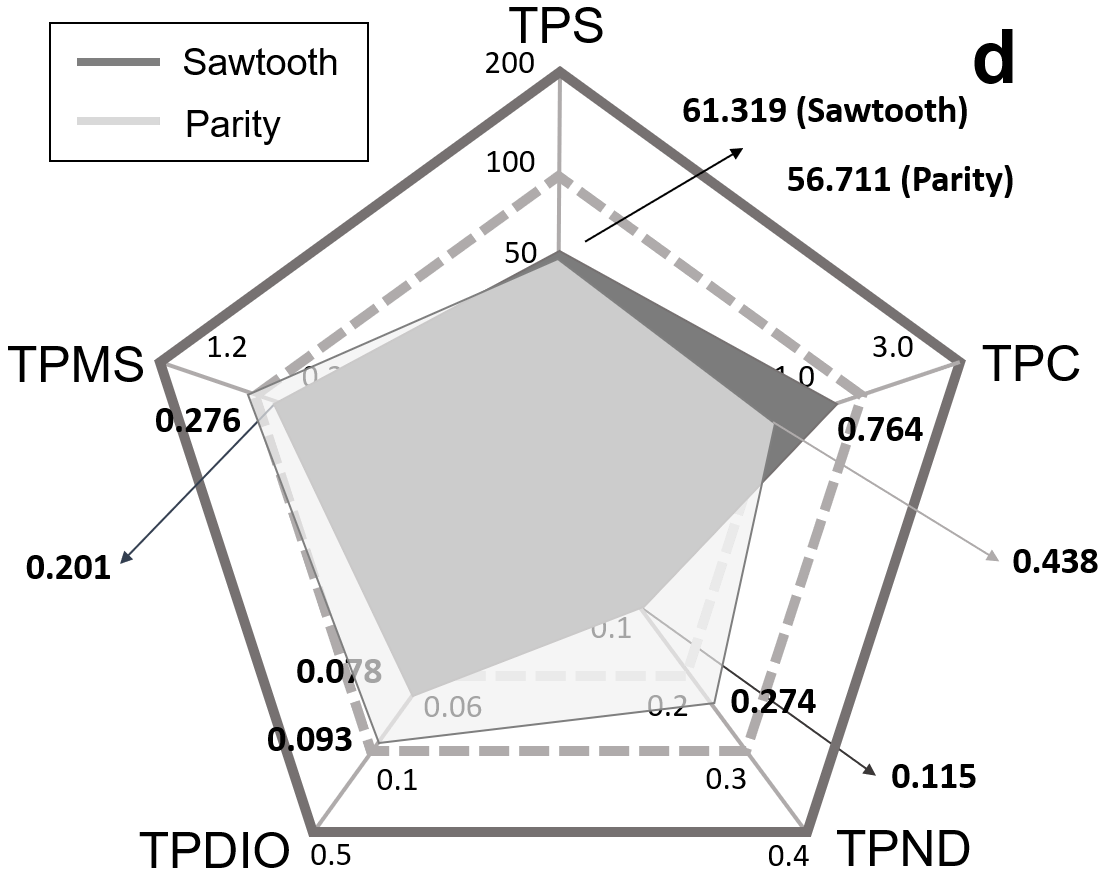}
	\end{minipage}
	\begin{minipage}{0.05\textwidth}
	\includegraphics[width=\textwidth]{fill.png}
	\end{minipage}
	\begin{minipage}{0.39\textwidth}
	\includegraphics[width=\textwidth]{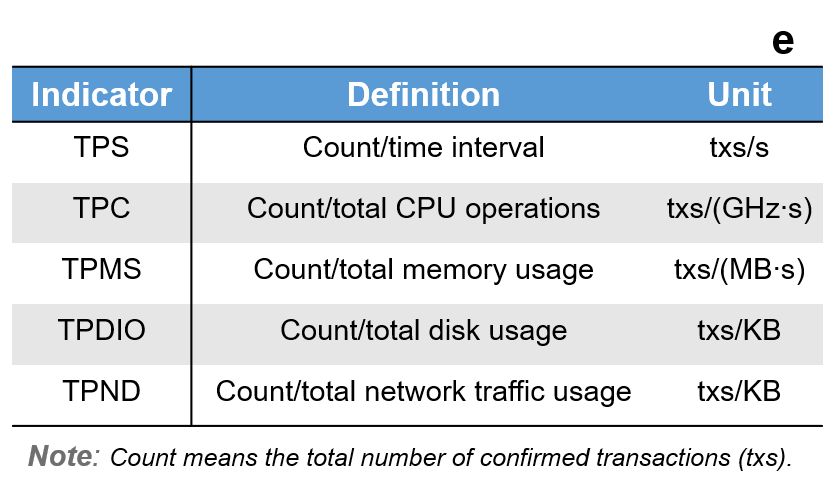}
	\end{minipage}
	\label{fig:computational-cost}
	\vspace{0.3cm}
	\caption{\textbf{Resource efficiency of classic consensus mechanisms.} \textbf{a-d,} \textit{Resource efficiency of PoW-based designs (\textbf{a}), BFT-based design (\textbf{b}), Graphene-based design (\textbf{c}) and other designs (\textbf{d})}. \textbf{e,} \textit{Definitions of the metrics adopted by \textsf{Prism}. We note that 5Mbps bandwidth is unable to deploy 10-node P2P network of Fabric v0.6 (resource-related benchmarking is conducted in the same testbed with performance benchmarking). Therefore, we reconfigure the testbed to LAN connections (up to 1000Mbps) when benchmarking Fabric v0.6 to avoid system crash. Such circumstance further sheds light on the TPND bottleneck of PBFT. TPC, Transactions per CPU; TPMS, Transactions per Memory Second; TPDIO, Transactions per I/O Disk; TPND, Transactions per Network Data. Data obtained by \textsf{Prism}.}}
\end{minipage}
\end{figure*}

\section{Resource Inefficiency behind The Scalability Constraints}\label{sec:resource efficiency}
Compared with centralized topologies, low resource efficiency attributed to the distributed computation and storage manners might be the "Achilles' heel" of blockchain. 
In order to quantitatively illustrate the efficiency of physical resources consumed by classic blockchain designs, we use \textsf{Prism} for conducting resource-related benchmarking. 
\textsf{Prism} adopts a systematic analyzing model (Fig. 4e), which defines resource efficiency as the number of transactions that could be confirmed consuming one unit of physical resource, including CPU operation, network traffic (upload and download), and storage space (memory and disk) \cite{8449244A}. 
By resource-related benchmarking, we expect to study why consensus-based scaling loses effectiveness and highlight the key factors affecting classic blockchain.
To ensure the coverage, for each category in Fig. 3, several representatives are selected and included in benchmarking scope.
Our observations will guide researchers to objectively evaluate the current blockchain techniques and explore the next-generation scaling strategies.

\textit{\textbf{Key resource factors.}} Fig. 4a-d illustrate the resource efficiency of 6 blockchain designs with 5 consensus mechanisms in detail. Overall, benchmarking on \textsf{Prism} well explains the reasons for the scalability deficiency in blockchain and is highly consistent with our previous observations. 

PoW-based designs are seriously trapped in poor TPC (Fig. 4a), because miners will perform hash collisions at the highest power when executing PoW, resulting in, on average, huge computation overhead. 
Moreover, the heavy mining processes occupy considerable RAM space, especially for the Ethash algorithm of Ethereum. 
In fact, Ethash frequently fetches the DAG dataset during mining.
Limited by the memory access time, the impact of computing power on mining is successfully reduced, known as memory-hard \cite{8695663A}. 
Although such design improves the network stability since computing power increases rapidly over time (Fig. 1a), the TPMS of Ethereum is decreased accordingly. 
More seriously, researchers fail to fundamentally improve the resource efficiency of PoW because the workload proof is still inevitable regardless of any evolution. 
TPC and TPMS will remain at low levels as long as mining exists.
In conclusion, although PoW-based designs acquire the capacity to accommodate large-scale networks, they can hardly be effectively scaled up.

PBFT-based Fabric v0.6 performs the lowest among all in terms of TPND (Fig. 4b). 
The stringent demand on network traffic and bandwidth severely impairs the scalability of PBFT-based designs. 
Fortunately, owing to the complex communication process among peers, CPU's burden acquires appreciably alleviated, enabling Fabric v0.6 to accommodate power-constrained devices, such as wearable devices and embedded sensors. 
In this case, network conditions, including scale, bandwidth, and stability, become the key factors determining blockchain's performance. 
With the development of 5G technology and network infrastructures, the application range of PBFT will gradually expand. 
Since Fabric v1.0 (in 2016), IBM no longer provided PBFT and replaced it with two lightweight ordering protocols, namely Solo and Kafka \cite{KAFKA}. 
Meanwhile, many researchers made further innovations based upon PBFT, reducing the original $O(n^2)$ traffic complexity to $O(n)$ without affecting the efficiency of other resources \cite{8490892A, Kapitza:2012:CRB:2168836.2168866A}. 
So, BFT-based blockchain will also support globally-distributed and large-scale deployments. However, considering that the workflow of BFT is well-studied, we opine that its potential has been exhausted and there exists narrow scaling space for BFT-based designs.

As illustrated in Fig. 4c, EOS shows comprehensive superiority of resource efficiency. 
During our benchmarking, CPU is occupied almost exclusively for OS and transaction processing, while the consensus workflows cause negligible CPU operations. 
Contributing to the minor memory occupancy and swaps required by Graphene, the TPMS and TPDIO of EOS are also higher than other blockchain designs. 
Note that EOS performs moderately with respect to TPND due to the fast block creations (6 blocks within 3s). 
Without exceeding the workload threshold, the block capacity is not thoroughly utilized --- many blocks are nearly empty. 
Interestingly, as the throughput increases, concurrent transactions will saturate all newly-created blocks, further exploiting block capacity and improving TPND. 
More importantly, similar to centralized projects, the hardware scaling of BPs directly promotes the performance of EOS network since the consensus workflow is thoroughly lightweight and the resource efficiency maintains invariant with varying resources. 
Nonetheless, considering that the hardware scaling is bounded by Moore's law, whose end is in sight, the scaling potential of Graphene also borders its upper bounds.

Apart from aforementioned categories, there are several distinctive consensus mechanisms drawing unique concepts and displaying individual features in resource efficiency. 
Well-known projects we measured are Proof-of-Authority \cite{878A} and Proof-of-Elapsed-Time (PoET) \cite{PoET}, whose results are illustrated in Fig. 4d. 
Clearly, their resource utilization is more balanced and stays at acceptable levels.

\begin{figure*}
\centering
\begin{minipage}{1\textwidth}
	\centering
	\begin{minipage}{0.25\textwidth}
	\includegraphics[width=1.01\textwidth]{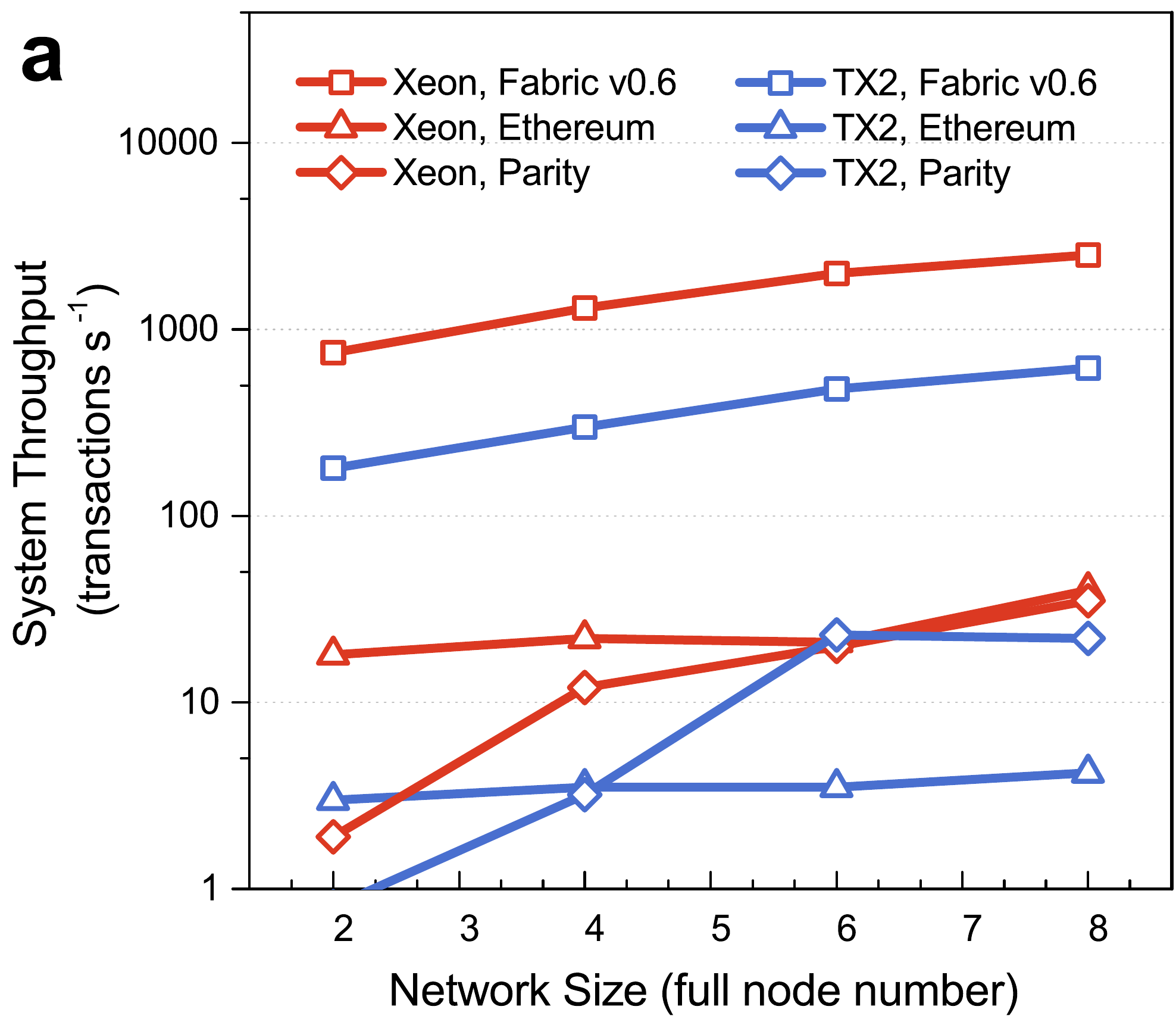}
	\end{minipage}
	\begin{minipage}{0.05\textwidth}
	\includegraphics[width=\textwidth]{fill.png}
	\end{minipage}
	\begin{minipage}{0.25\textwidth}
	\includegraphics[width=0.973\textwidth]{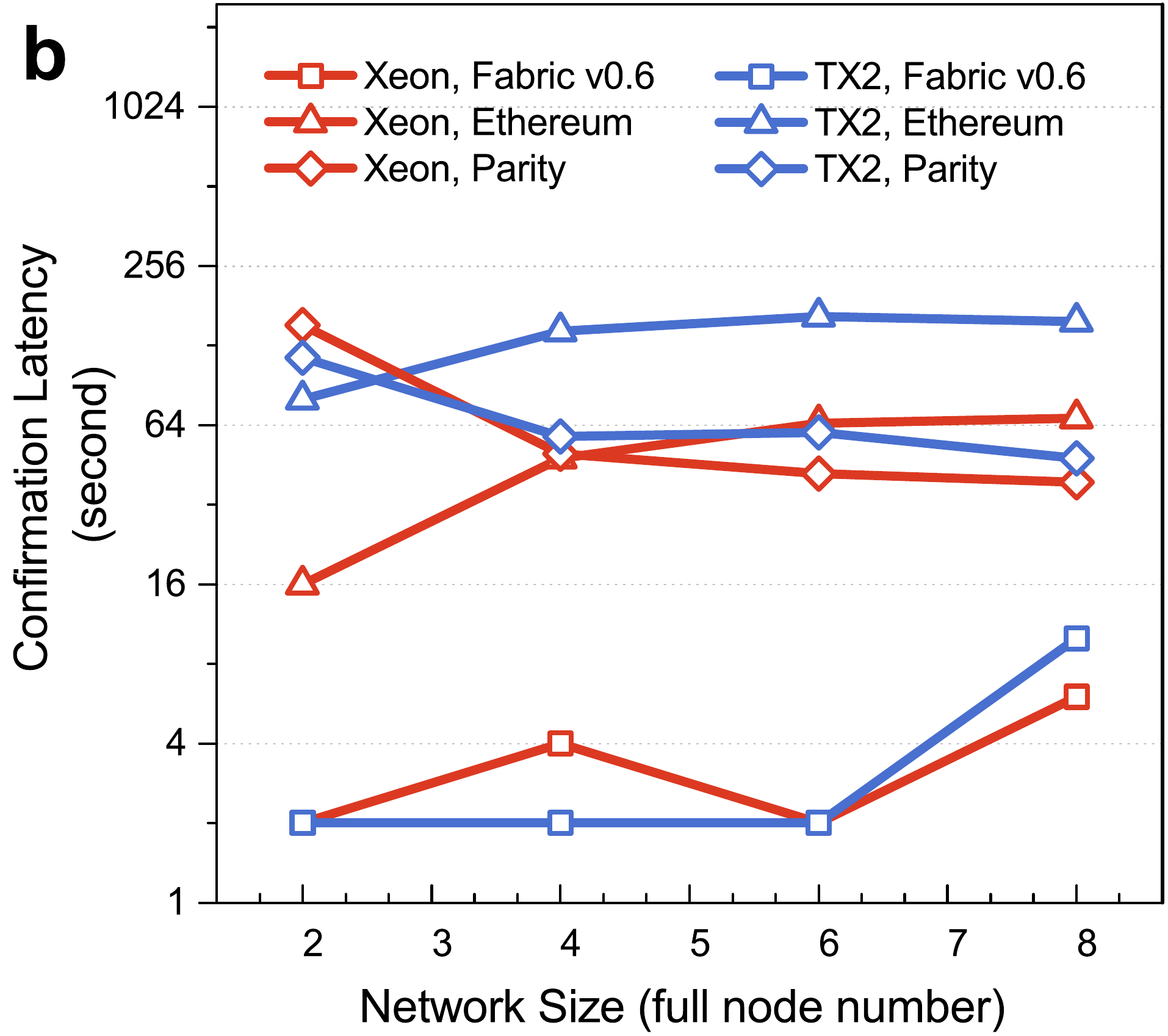}
	\end{minipage}
	\begin{minipage}{0.04\textwidth}
	\includegraphics[width=\textwidth]{fill.png}
	\end{minipage}
	\begin{minipage}{0.25\textwidth}
	\includegraphics[width=0.98\textwidth]{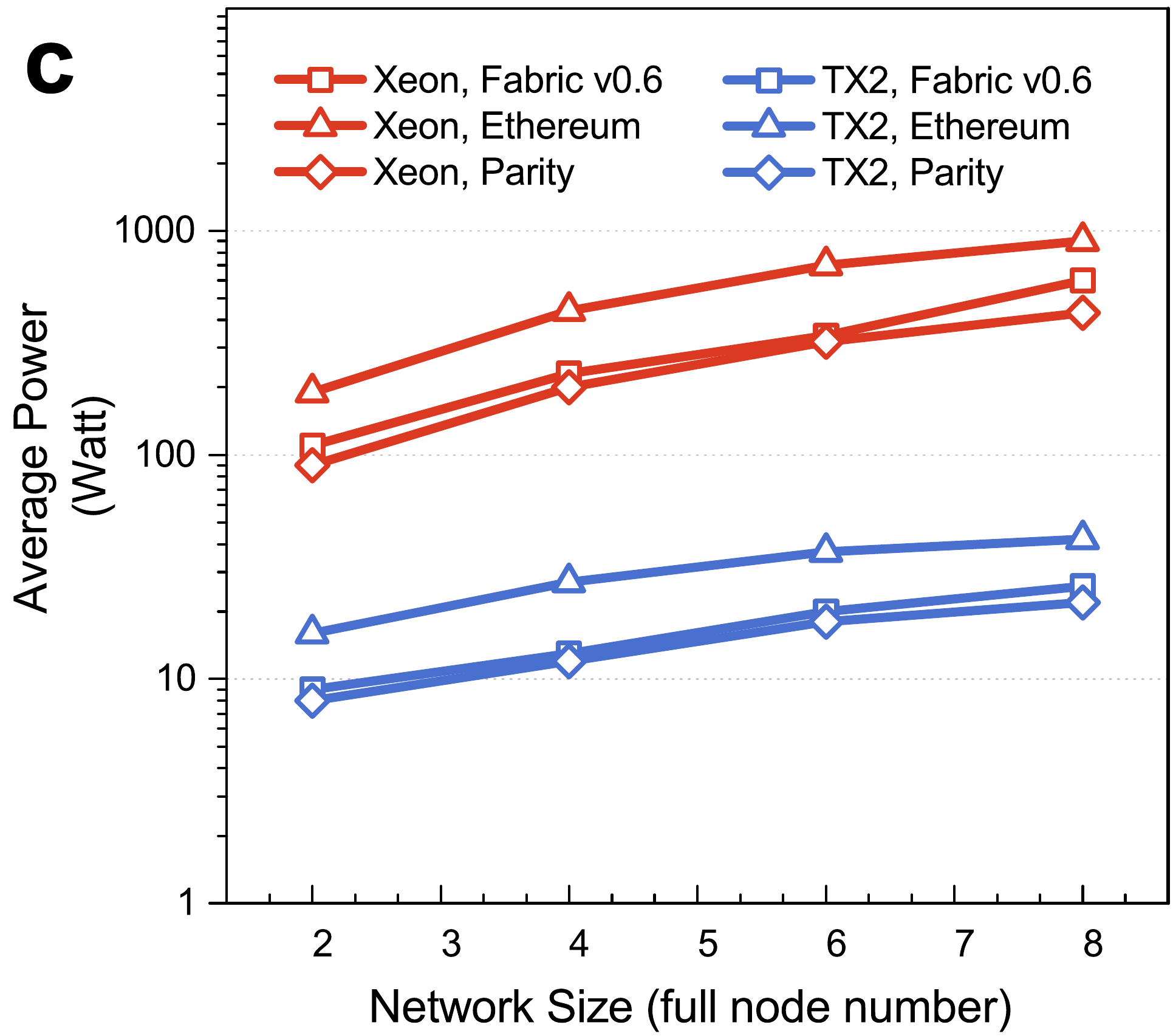}
	\end{minipage}
	\label{fig:computational-cost}
\end{minipage}
\end{figure*}
\begin{figure*}[htbp]
\vspace{-0.1cm}
\centering
\begin{minipage}{1.001\textwidth}
\centering
	\begin{minipage}{0.54\textwidth}
	\includegraphics[width=\textwidth]{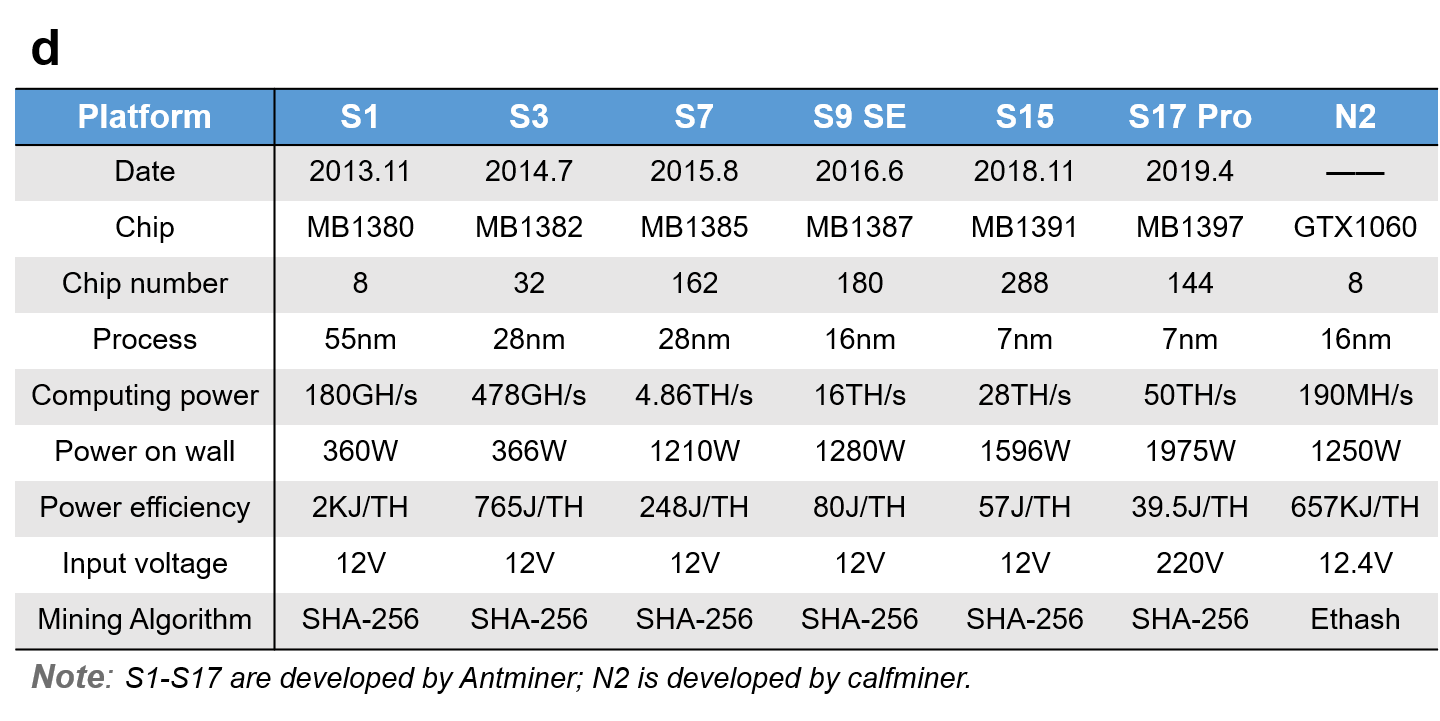}
	\end{minipage}
	\begin{minipage}{0.04\textwidth}
	\includegraphics[width=\textwidth]{fill.png}
	\end{minipage}
	\begin{minipage}{0.28\textwidth}
	\includegraphics[width=\textwidth]{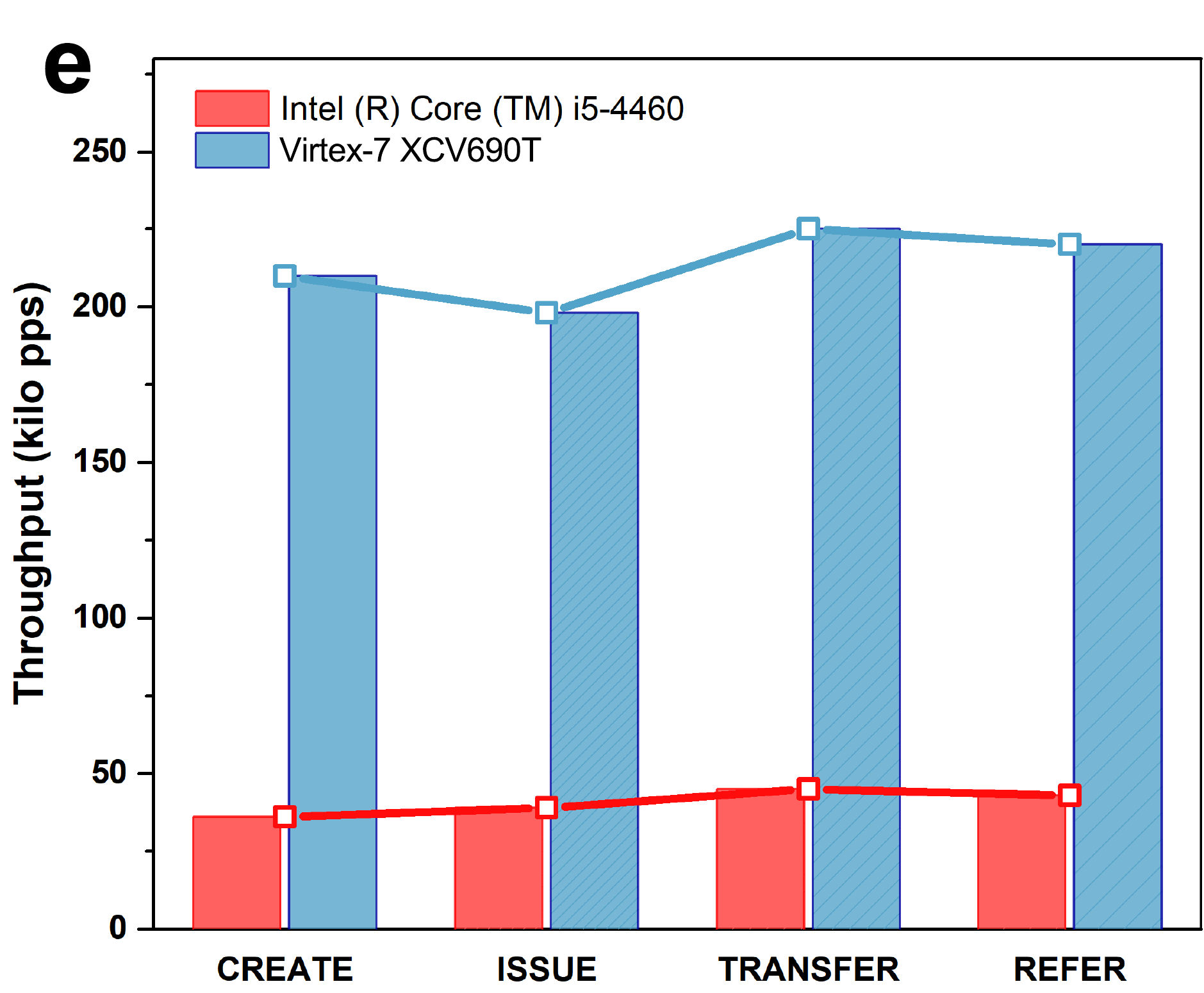}
	\end{minipage}
\end{minipage}
\end{figure*}
\begin{figure*}[htbp]
\vspace{-0.1cm}
\centering
\begin{minipage}{1.001\textwidth}
\centering
	\begin{minipage}{0.6\textwidth}
	\includegraphics[width=\textwidth]{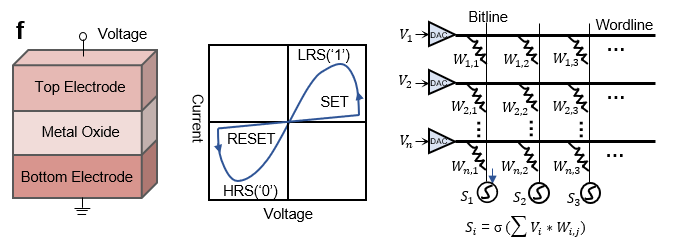}
	\end{minipage}
	\begin{minipage}{0.003\textwidth}
	\includegraphics[width=\textwidth]{fill.png}
	\end{minipage}
	\begin{minipage}{0.37\textwidth}
	\includegraphics[width=\textwidth]{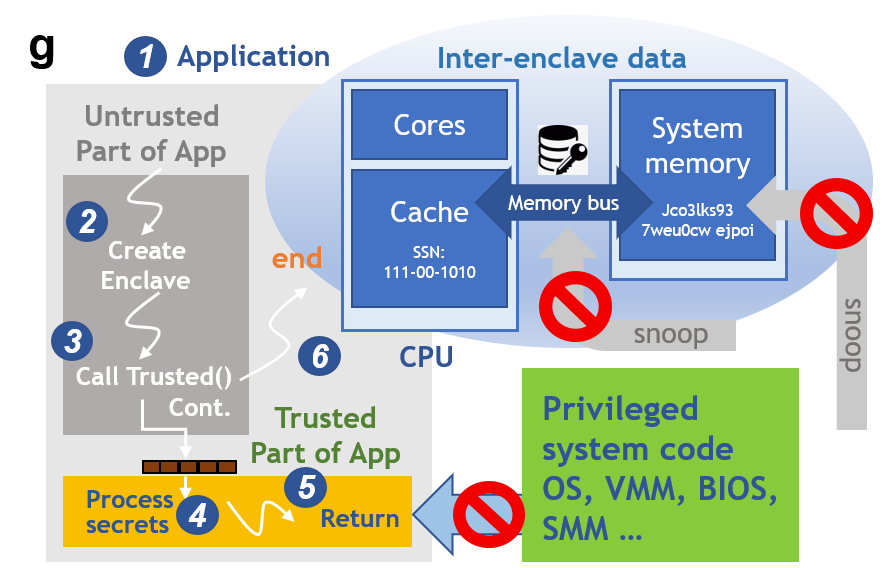}
	\end{minipage}
	\label{fig:computational-cost}
	\caption{\textbf{Hardware assistance in blockchain.} \textbf{a-b,} \textit{Correlations of the hardware configurations and blockchain performance on throughput (\textbf{a}) and confirmation latency (\textbf{b}).} \textbf{c,} \textit{Average power consumption of brawny and wimpy cluster}. \textbf{d,} \textit{Detailed parameters of advanced miners published in 2013-2019.} \textbf{e,} \textit{FPGA accelerations of network commands for blockchain.} \textbf{f,} \textit{Basics of ReRAM. As a in-memory conputing device, ReRAM cell contains two eletrodes and a metal-oxide layer. Applying an external voltage across ReRAM, it can be switched between HRS (logic "1") to LRS (logic "0"). A series of cells can construt a crossbar connected by bitlines and support blockchain operations}. \textbf{g,} \textit{Workflows of SGX-protected applications. Steps: 1) application builds trusted and untrusted parts; 2) application creates the enclave (in trusted memory); 3) trusted function is called, execution transitions to the enclave; 4) enclave controls all process data and denies external access; 5) trusted function returns, enclave data remains intact; 6) application continues normal execution. VMM, Virtual Machine Monitor; BIOS, Basic Input Output System; SMM, System Management Mode. Credit: panel a-c adapted from \cite{ENERGY1}, IEEE; panel e adapted from \cite{8672299A}, IEEE; panel f adapted from \cite{8942057}, IEEE; panel g adapted from \cite{SGX}, Intel Inc.}} 
\end{minipage}
\end{figure*}

\textit{\textbf{Overcoming resource inefficiency.}} After sorting out these statistics, we find that the potential of consensus-based scaling is limited, as the resource efficiency of the successively proposed consensus mechanisms still holds on the same order of magnitude. 
Even sacrificing the decentralization, this metric obtains no decisive rise in Graphene. 
In addition, the resource efficiency we benchmark, denoted by $R$, is targeted at one single peer. 
Provided that distributed manners are adopted, the resource consumption of \textit{n}-node P2P network roughly surges to $n \times R$, so the resource efficiency decreases to $R/n$. 
In short, blockchain's resource efficiency is negatively correlated with the network scale. 
Therefore, topology-based scaling is essential to scale up blockchain since multi-chain techniques can optimize the resource efficiency of $n$-node P2P network to $(R\,\times\, c)\,/\,n$, where $c$ indicates the amount of entities that independently allocate resources. 
By adjusting $c$ to near $n$, blockchain obtains the capability of maintaining the resource efficiency unchanged under any network scale. 
Finally, the performance of the entire blockchain network will grow with more peers joining in, which agrees with our ultimate objective of effectively scaling up blockchain.

\section{Hardware assistance in blockchain}\label{sec:hardware assistance in blockchain}
From the perspective of hardware, blockchain networks act as distributed computers whose resources are aggregated by scattered peers.
As depicted in Fig. 4, blockchain consumes computation resources for consensus, network traffic for communication, RAM for executing smart contracts, and storage room for saving ledger states.
In the past few decades, Moore's law-based complementary metal-oxide semiconductor (CMOS) scaling empowered integrated circuits to cram more components, resulting in ever-increasing performance density and efficiency \cite{CMOSA}. 
Advanced devices continuously enter in blockchain networks, enlarge the resource volume, and thus bring either positive or negative effects on scalability (Fig. 1a).
Here, the hardware assistance in blockchain is summarized with the following three strategies.

First, the hardware advances can directly scale up the blockchain. 
Intuitively, for high-end computers, the performance at which they run blockchain no doubt outperforms that of deploying blockchain on low-end ones, the same for distributed computer clusters (that is, P2P networks). 
Recall that for EOS, the computing power, bandwidth, and power supplement of BPs determine the capability of the whole network. 
To this end, candidates owning high-end servers are favored during BP elections. 
Apart from EOS, Fig. 5a-b illustrate the performance of three blockchain designs in two homogeneous clusters, where high- and low-end devices represent Intel Xeon \cite{Xeon} and NIVIDIA TX2 \cite{TX2}, respectively. 
The comparison apparently proves a positive correlation between the performance of hardware and that of blockchain. 
However, researchers might get impeded when attempting at the further scaling of blockchain devices in the future, given the end of Moore's law is in sight (the latest CMOS chips have reached 7nm, Moore's law will finish at around 5nm \cite{Moore}).
The anticipated ending of CMOS scaling further highlights the scalability bottleneck of blockchain and makes the desire for next-generation scaling strategies more urgent.

Despite the higher performance, adopting advanced devices with faster execution speed will increase the energy overhead accordingly (Fig. 5c). 
Previous work presented that wimpy networks with balanced performance-to-power ratios can achieve reasonable performance while saving much energy, which is a meaningful trade-off for applications \cite{ENERGY1}.

Second, suitable hardware architectures can coordinate with specific mining algorithms to maximize miners' economic profit \cite{7383912A, 4560238A, 8741638A, 7440392A, 8698069A, 8783449A}. 
Naturally, this strategy is aimed at PoW-based blockchain.
Taking Bitcoin as an example, during 2009-2019, the equipment of Bitcoin mining experienced multiple rounds of revolution \cite{8048662}. 
Initially, Bitcoin miners used CPUs since they are widely-adopted and general-purpose processors. 
However, CPUs pursue versatility by integrating massive units for handling different tasks. 
A considerable proportion of such units contribute little to mining, for instance, memory, cache, and control units. 
Conversely, due to the scanty amount of arithmetic and logic units (ALUs), CPUs lack the capability of executing parallel computing, which is requisite for operating SHA-256 --- Bitcoin's mining algorithm. 
Equipped with numerous stream processors and cores, GPUs rapidly overtook CPUs as the first choice of Bitcoin miners. 
Around 2013, miners found that ASIC chips can greatly enhance the computing power via customized design and stacking more ALUs \cite{4415766A, 8701906A, 5937967A}. 
Hence, ASICs further surpassed GPUs and have became prevalent in Bitcoin mining industry nowadays, as shown in Fig. 1a. 

Similar to Bitcoin, other blockchain can exploit the architectural features of certain hardware as well. 
For instance, Ethereum's Ethash and IOTA's Curl are suitable for GPUs, while mining under Storj relies on the hard drive \cite{Storj}. 
The parameters of the most advanced mining equipment in each era are described in Fig. 5d. 
In general, the hardware updates motivated by the pursuit of faster mining is the most attractive aspect of hardware applications in blockchain.
Unfortunately, such updates merely serve the miners, while making tiny technical assistance in the scalability optimizations.
On the contrary, the computing power competition become fierce.
Power-constrained peers, whose devices lag behind those of rich "competitors", are expelled from mining.
In the long term, the gaps among participants will expand in degree, intensifying the Matthew Effect \cite{ME}.
With the end of Moore's law, quantum computing technology is entering the blockchain industry.
Quantum computers will produce profound impact on blockchain encryption, mining, etc \cite{QC}.

Finally, the customized and trusted computing structures can accelerate certain blockchain operations which are time-consuming or resource-intensive. 
Naturally, CPUs already satisfy all functional requirements of blockchain. 
However, for some relatively complex operations, such as caching, data transferring, and transaction validation, specifically designed hardware sometimes outperforms ordinary CPUs in reducing the execution time (Fig. 5e). 
The most promising accelerators for blockchain are FPGA, GPU, and ReRAM (Fig. 5f) \cite{8942056, 8942057}.
Since the time-consuming tasks are handled by accelerators, blockchain's performance attains significant improvement. 
In 2013, Intel developed the Software Guard Extensions (SGX) technology, which built trusted execute environments (TEE) in physical memory, called enclaves (Fig. 5g) \cite{SGX}. 
Currently, SGX technology is compatible with multitudes of CPU series, like Intel Xeon, Skylake, and Core. 
Protected by SGX, even the privileged malware at the OS, BIOS, VMM, or SMM layers cannot access the content of local enclaves, provisioning a great simplification possibility for consensus workflows. 
Since enclaves can be viewed as trusted, SGX successfully acts as CAs to promote the resource efficiency without undermining network decentralization. 
We suggest that with such hardware-assisted scaling, blockchain can find a simple approach towards improving scalability. 

\vspace{-0.1cm}
\section{Towards the Next-Generation Scaling}
Although consensus-based scaling and Moore's law accompanied blockchain's leap from 1.0 era (cryptocurrencies) to 3.0 era (decentralized applications) and witnessed its great prosperity in countless fields, simply modifying consensus or stacking resources can never achieve satisfying scalability. 
To provision new impetus for blockchain techniques, researchers are attempting to conceive the next-generation scaling strategies, especially the topology-based scaling from network architects and the hardware-assisted scaling from engineers.

\subsection{Topology-Based Scaling}
\textit{\textbf{Offloading workloads.}} Off-chain is a classic technique which enhances network concurrency by constructing channels \cite{ONOFFB}.
In traditional on-chain topologies, myriad transaction confirmations concentrate on the blockchain network with limited capacity.
Off-chain networks, in contrast, build independent and long-lived channels over which an arbitrary amount of transactions can be processed locally between individuals.
The blockchain is contacted only when peers initialize and close channels, or the involved parties encounter conflicts, in which case the preset contracts will handle fair settlement.
This topology therefore effectively offloads the vast majority of blockchain network's workloads, accommodating more channels and peers.
Simultaneously, both sides of channels are free of synchronizing network-wide message or participating in consensus, so the resource efficiency and performance at the application layer are significantly improved.
Viewed from blockchain network, however, the resource efficiency keeps unchanged because of no architecture innovations.

The well-known Bitcoin Lightning network \cite{LIGHTNET1} and duplex micropayment channel \cite{8695663411A} first explored off-chain to overcome the scalability bottleneck.
Furthermore, Dziembowski \textit{et al.} \cite{8835315B} presented the concept of virtual payment channel, which introduced smart contracts to bypass undependable intermediaries when processing cross-channel payments.
In order to support general applications rather than cryptocurrencies, they further polished their proposal by building channels recursively and successfully constructed the state channel allowing arbitrarily complex smart contracts to execute \cite{Dziembowski:2018:GSC:3243734.3243856B}.
All these designs achieved promising performance and a potential far higher than consensus-based scaling.
Moreover, the security of off-chain technology has been elaborated in detail \cite{Khalil:2017:RRO:3133956.3134033B, Malavolta:2017:CPP:3133956.3134096B}.
Some researchers even claimed that their proposals can sustain dependability when the peers at both sides of one channel are malicious.

\textit{\textbf{Enabling interoperability.}} Parallel-chain techniques, including side-chain, cross-chain, and child-chain, intend to enable the interoperability among multiple blockchains.
This concept was first proposed by Back \textit{et al.} \cite{SIDECHAIN}, whose original goal was to build isolated side-chains for conducting low-cost experiments since the risk of arbitrary optimizations on the parent-chain (in \cite{SIDECHAIN}, Bitcoin) is unbearable.
Both side- and parent-chain are completely independent, which means they are maintained by different sets of peers.
Through cross-chain communication mechanisms, like two-way peg protocols, numerous side-chains can be interconnected and allow for secure data transfers.
From the perspective of transaction processing, parallel-chain partitions the workloads that previously belong to one single chain \cite{8431965B}.
Suppose that each side-chain could process $\varphi$ transactions within 1s, the throughput of the whole blockchain network containing $\omega$ parallel chains reaches $\varphi\, \times \,\omega$ TPS, immensely lifting up the scalability. 
The child-chain is a hierarchical form of parallel-chain wherein various child-chains process the concurrent transactions, while the parent-chain completes the confirmations by verifying Merkle Tree roots \cite{childchain}.

Nano \cite{nano} and Chainweb \cite{chainweb} explored parallel-chain in the industrial and academic fields, both of which showed outstanding superiority over traditional single-chain systems, especially in throughput.
Going forward, parallel-chain could switch the consensus mechanisms to adapt to different application scenarios, for example, adopting PoW for payment platforms and PoET for supply chains.
As a result, the flexibility of blockchain likewise experiences significant growth.
Since the parallel-chain scaling avoids altering the consensus workflows of the parent-chain, it is particularly suitable for the projects which have been put into widespread practical deployments.
Considering that such projects have attracted considerable capital, any scaling attempt should be handled very conservatively.

\textit{\textbf{Partitioning network.}} After 2016, the combinations of sharding techniques with blockchain have received in-depth research.
The representative designs include NUS's ELASTICO \cite{Luu:2016:SSP:2976749.2978389B}, EPFL's OmniLedger \cite{8418625B}, Visa's RapidChain \cite{Zamani:2018:RSB:3243734.3243853B}, and NUS's TEE-assisted sharding system supporting general applications \cite{Dang:2019:TSB:3299869.3319889B}.
In blockchain, sharding-based scaling is realized by partitioning or parallelizing the heavy workloads among multiple smaller committees (shards).
Accordingly, the basic unit of consensus and resource allocation changes to shards, where a reasonably small number of interconnected peers merely synchronize and store data within their own shard.
Each shard can be regarded as an independent and fully functional blockchain, which processes disjoint sets of transactions without supervision from any trusted administrators.
Thus, either for a single peer or the entire P2P network, the resource efficiency is maximized.
An important breakthrough made through full sharding is that blockchain first attains the capability of linearly improving performance with the rising number of peers, thereby achieving an equivalent scalability as Visa.

Innovations of sharding designs concentrate on three key mechanisms: shard formation (initialization and reconfiguration), intra-shard consensus, and cross-shard communication.
Since shard formation directly determines the intra-shard security, for example, the proportion of malicious peers in the newly generated shards, this process is required to be bias-resistant.
Otherwise, attackers would get a great opportunity to gather in the same shard. 
ELASTICO \cite{Luu:2016:SSP:2976749.2978389B}, the first sharding-driven blockchain, leveraged PoW-based seeds as the randomness to assign peers into corresponding shards, which was proved to have hidden dangers.
Motivated by this, Kokoris-Kogias \textit{et al.} \cite{8418625B}, Dang \textit{et al.} \cite{Dang:2019:TSB:3299869.3319889B}, and Zamani \textit{et al.} \cite{Zamani:2018:RSB:3243734.3243853B} proposed various randomness generation mechanisms based on RandHound protocol, TEE, and verifiable secret sharing protocol, respectively, to ensure the complete resistance towards bias.

Moreover, the intra-shard security decays over time because attackers tend to move towards those vulnerable shards, especially for the designs whose shard size is small.
In order to hold high-security, sharding systems have to frequently reassign peers, named reconfiguration, which is undoubtedly unfavorable.
Extending the reconfiguration interval needs to be specifically designed according to different intra-shard consensus.
For PoW-driven shards, the Chu-ko-nu mining presented in Monoxide weakened the influence of high-end peers by diluting their computing power to multiple shards without affecting the resource efficiency \cite{227661B}.
Similarly, RapidChain refined the synchronous consensus protocol of Ren \textit{et al.} and enhanced the robustness of BFT-driven shards \cite{8525387B}.
PoW and BFT are the intra-shard consensus mechanisms commonly used in recent sharding designs.
This is because they are well-proven and display excellent stability, and can thus be deployed in numerous parallel shards.

Apart from shard formation and intra-shard consensus, the last concern is about cross-shard communication.
In sharding designs, atomicity, isolation, and liveness are indispensable properties for securely processing pending transactions whose inputs/outputs belong to multiple shards. 
Atomicity demands that cross-shard transactions ought to be either confirmed or aborted by all shards, otherwise the funds will be persistently locked.
Similarly, isolation is defined as the independence and serializability of validating each cross-shard transaction owning multiple input/outputs.
As the first work that achieves atomicity, OmniLedger used active clients as the coordinators between input- and output-shards.
However, if the delegated clients are malicious, OmniLedger will fall into the denial of service and fail to process pending transactions within a finite latency \cite{Dang:2019:TSB:3299869.3319889B}.
Liveness just refers to the systems' resistance towards such conditions.
Fortunately, by combining two-phase protocols with BFT, Dang \textit{et al.} \cite{Dang:2019:TSB:3299869.3319889B} recently have addressed the concerns about isolation and liveness.

\textit{\textbf{Trade-offs.}} Despite the enormous potential displayed from topology-based scaling, two trade-offs are yet to be judged.
Recall that this approach scales up traditional single-chain designs via building channels, interconnecting isolated blockchains, or partitioning networks; the first trade-off is between the interoperability and complexity attributed to cross-chain/channel communication.
After topology-based scaling, in most cases, the "one chain, one asset" maxim of blockchain industry will be destroyed \cite{SIDECHAIN}.
Meanwhile, advanced topologies inevitably produce extra resource overhead.
For example, Back \textit{et al.} \cite{SIDECHAIN} adopted complicated two-phase protocols for secure transfers among heterogeneous parallel-chain, which might offset the increase in resource efficiency.
Nonetheless, the overall benefits of cross-chain communication outweigh their potential risks, let alone the central role of interoperability for the future blockchain applications \cite{8416383B}.
Therefore, this trade-off does not seem to be so confusing.

However, the second trade-off exists between the increased scalability and the decline in security and decentralization, as noted earlier, the "trilemma".
In view of the finiteness of available resources and their growth rates owned by blockchain networks, simultaneously improving scalability, decentralization, and security is impossible.
When developing Bitcoin, Nakamoto gave the top priority to absolute security, decentralization, and the robustness to accommodate thousands of peers, while he temporarily shelved the scalability by choosing a resource-intensive, low-performance PoW.
The superb security just explains why PoW could support more than 90\% of the worldwide digital assets [62].
Graphene, in contrast, attains a performance leap with the assistance of authoritative BPs; thus, EOS network exhibits a certain degree of partial decentralization.

Although researchers repeatedly emphasized their efforts towards holding security and decentralization, regardless of off-chain, parallel-chain, or sharding, most topology-based scaling designs still employed certain authoritative entities to empower some complex mechanisms.
These entities include off-chain channels, final committee in ELASTICO, reference committee in RapidChain, and clients in OmniLedger.
Additionally, the multi-chain networks after scaling are inevitably not as resistant to attackers as their prototypes, wherein all available resources are invested to protect only one chain.
Nevertheless, if we arbitrarily refuse topology-based scaling because of several flaws, blockchain might sidestep a perfect chance for scaling from the order of magnitude.

Envisioning the potential of topology-based scaling, we advise that users could accept a slight attenuation of decentralization to embrace high-quality blockchain services, as Bitcoin also encounters similar issues from the Matthew effect.
Meanwhile, researchers should keep on innovating to guarantee the security of multi-chain networks, eliminate the dependence on authoritative entities, and simplify cross-chain communication.

\vspace{-0.1cm}
\subsection{Hardware-Assisted Scaling}
\textit{\textbf{Accelerating time-consuming executions.}} Observing the intensive time and resource consumption of blockchain, many works have tried to accelerate or lighten certain operations being frequently performed.
Although fundamentally overcoming the scalability bottleneck seems impossible, hardware-assisted scaling excels in feasibility because it mainly rebuilds the application or consensus layer, while retaining the network topology.
In effect, customized FPGAs have attracted widespread interests for accelerating neural networks, machine learning, and blockchain \cite{8574545A, 8721457A}.

In most cases, FPGA accelerators are designed based on FPGA Network Interface Card (NIC) and connected to the host CPU (full nodes) via Peripheral Component Interconnect Express (PCIe). 
Since FPGA integrates NIC, RAM modules, and ALUs in one chip, the resource efficiency is much higher than that of standard CPUs.
Furthermore, compared with CPU, FPGA can process more requests in one batch because of its stronger capability in parallel computing, which is vital for handling massive threads. 
Recent work has shown that employing FPGA accelerators instead of host CPUs for responding to outside requests can effectively offload the workloads for peers.
Sakakibara \textit{et al.} \cite{8672299A} proposed an FPGA-based NIC, which collaborated with off-chain channels in rapidly transferring digital assets without interacting with blockchains.
Sanka \textit{et al.} \cite{8638204A} designed an FPGA platform integrating hash tables and several hash cores for caching Key-Value Store (KVS)-styled data, achieving a 103-fold improvement in response speed over CPU storage. 
With the assistance from FPGA, full nodes communicate with clients only when the required data or commands are not satisfied by accelerators, thereby saving considerable resources for executing blockchain operations.
As to outside clients, the throughput and latency of interacting with full nodes are optimized correspondingly, that is, a higher quality of service.
Apart from FPGA, other hardware can also serve as blockchain accelerators. 
For example, Morishima \textit{et al.} \cite{8374464} applied GPUs in the local Patricia tree-structured KVS memory to reduce the time of searching metadata.
Note that the KVS which is widely adopted in blockchain can combine hash tables or be organized as tree structures to adapt to FPGAs and GPUs, thereby improving the speed of data access \cite{6645520A, Hetherington:2015:MSS:2806777.2806836A}.
Luo \textit{et al.} \cite{8203789} presented a tracetrack memory based in-memory computing scheme for conducting encryption/decryption in blockchain, which achieved 8 folds of throughput per area compared with the state-of-the-art CMOS-based implementation.
As to ASICs, considering the high development cost, they are temporarily unsuitable to accelerate blockchain operations except mining.

\textit{\textbf{Lightening resource-intensive mechanisms.}} PBFT and its variants achieve high throughput (ideally, more than 10000 TPS) since the computation-intensive mining is abolished.
Instead, the three-phase multicast is used to atomically exchange messages, which leads to a communication complexity of $O(n^2)$ and cannot accommodate enough peers of public-chain applications.
Therefore, researchers have long tried to perfect BFT and later realized that a feasible approach is to apply a hybrid fault model --- some components are trusted and only fail by crashing, while the others remain BFT and are possible to launch any malicious attacks.
Following this strategy, the three-phase multicast protocols could be lightened into two-phase ones, thereby eliminating the bottleneck on TPND that hinders BFT.
Note that we exclude the hybrid fault model from consensus-based scaling because we merely discuss pure fault model in that part.
The premise of hybrid fault model is to ensure the security and validity of trusted subsystems, so it calls for the assistance from trusted hardware technique.
As early as 2004, Correia \textit{et al.} \cite{1180168A} applied Trusted Timely Computing Base (TTCB), the tamper-proof and interconnected components distributed in host CPUs, to provide trusted ordering services for BFT.
Chun \textit{et al.} \cite{Chun:2007:AAM:1294261.1294280A} presented two consensus mechanisms based on Attested Append-Only Memory (A2M), building a trusted log that makes any tampering attempts detectable.

However, A2M needs to store all certified messages as a part of the subsystem --- a prominent defect \cite{Behl:2017:HSS:3064176.3064213A}.
The principal goal behind hybrid fault models is to control the size and complexity of protected components since the larger the trusted code is, the more the potential vulnerabilities expose.
Following this principle, Kapitza \textit{et al.} \cite{Kapitza:2012:CRB:2168836.2168866A} implemented a small trusted component on FPGA-based platform, for improving the security and saving resources.
MinBFT \cite{6081855A} and TrInc \cite{Levin:2009:TST:1558977.1558978A} employed trusted counters for attestations or to associate sequence numbers to each operation, making the protocol simpler and easier to deploy.
TEE technique, for instance, Intel SGX, ARM TrustZone, and Sanctum has also been explored to create trusted subsystems, and gradually become prevalent. 
Behl \textit{et al.} \cite{Behl:2017:HSS:3064176.3064213A} proposed a highly parallelizable and formal hybrid consensus mechanism named Hybster, which used SGX to multiply the trusted subsystems.
Dang \textit{et al.} \cite{Dang:2019:TSB:3299869.3319889B} installed A2M in SGX for optimizing PBFT and generated bias-resistant randomness for shard formations.
Moreover, Liu \textit{et al.} \cite{8419336A} developed a resource aggregation algorithm that combined TEE with lightweight secret sharing to reduce the communication complexity to $O(n)$. 

These above works can be seen as the forerunners of the scaling attempts which no longer rely on refining consensus, while aiming to consume resource in a more efficient manner. 
Extensive experiments and deductions demonstrate the potentials of such proposals. 
However, they are still in an early stage of development since the statistics are merely staying in laboratories or private benchmarks.

\section{Conclusion}
This Perspective discusses the major concerns about the scalability of blockchain technique. 
Specifically, we suggest that the consensus innovations and Moore's law-based hardware scaling, which were widely adopted in the past decade, no longer provides power for blockchain to satisfy application requirements.
Based on in-depth benchmarking and analysis, we find the hidden factor under blockchain's scalability constraints, that is, the resource inefficiency.
We also present a vision for emerging strategies that can effectively scale up blockchain in the future, including topology-based scaling and hardware-assisted scaling.
\vspace{0.03cm}
\normalem
\bibliographystyle{IEEEtran}
\bibliography{IEEEabrv,icc}

\end{document}